\documentclass[10pt]{article}

\usepackage{epsfig} 
\usepackage{geometry}                                                           
\usepackage{graphicx}
\usepackage{dcolumn}
\usepackage{color}
\usepackage{bm}
\usepackage{subfigure}

\begin{document}

{\LARGE \noindent \bf Nonmonotonic particle-size-dependence of magnetoelectric coupling in strained nanosized particles of BiFeO$_3$ } \footnote{Correspondence and requests for materials should be addressed to S.G. (drsudiptagoswami@gmail.com) and D.B. (dipten@cgcri.res.in)}

\vspace{5mm}

\noindent Sudipta Goswami$^1$, Dipten Bhattacharya$^2$, Chandan K. Ghosh$^3$, Barnali Ghosh$^4$, S.D. Kaushik$^5$, V. Siruguri$^5$ and P.S.R. Krishna$^6$

\vspace{10mm}

{\small \noindent $^1$Department of Solid State Physics, Indian Association for the Cultivation of Science, Kolkata 700032, India, $^2$Nanostructured Materials Division, CSIR-Central Glass and Ceramic Research Institute, Kolkata 700032, India, $^3$School of Materials Science and Nanotechnology, Jadavpur University, Kolkata 700032, India, $^4$Department of Materials Science, S. N. Bose National Center for Basic Sciences, Kolkata 700098, India, $^5$UGC-DAE Consortium for Scientific Research, Bhabha Atomic Research Centre, Mumbai 400085, India, $^6$Solid State Physics Division, Bhabha Atomic Research Centre, Mumbai 400085, India}

\date{\today}

\begin{abstract}
Using high resolution powder x-ray and neutron diffraction experiments, we determined the off-centered displacement of the ions within a unit cell and magnetoelectric coupling in nanoscale BiFeO$_3$ ($\approx$20-200 nm). We found that both the off-centered displacement of the ions and magnetoelectric coupling exhibit nonmonotonic variation with particle size. They increase as the particle size reduces from bulk and reach maximum around 30 nm. With further decrease in particle size, they decrease precipitously. The magnetoelectric coupling is determined by the anomaly in off-centering of ions around the magnetic transition temperature ($T_N$). The ions, in fact, exhibit large anomalous displacement around the $T_N$ which is analyzed using group theoretical approach. It underlies the nonmonotonic particle-size-dependence of off-centre displacement of ions and magnetoelectric coupling. The nonmonotonic variation of magnetoelectric coupling with particle size is further verified by direct electrical measurement of remanent ferroelectric hysteresis loops at room temperature under zero and $\sim$20 kOe magnetic field. Competition between enhanced lattice strain and compressive pressure appears to be causing the nonmonotonic particle-size-dependence of off-centre displacement while coupling between piezo and magnetostriction leads to nonmonotonicity in the variation of magnetoelectric coupling.
\end{abstract}

\vspace{5mm}

{\Large \noindent \textbf{Introduction}} 

\noindent Observing different patterns of anomalous ion movement around the phase transition in a solid is rewarding as it generates deeper insight about the nature of phase transition \cite{Calamiotou}. For example, by recording the positional shift of the oxygen ions in amorphous ice around a first-order-like phase transition under pressure, it is possible to see how next-nearest-neighbor ions undergo static displacement beyond a critical pressure to fill the empty interstitials partially while the first-nearest-neighbor ions maintain their positions intact throughout \cite{Klein}. This result helps in explaining the observed sharp rise in density of the amorphous ice above critical pressure. The anomalous ion movement near the phase transition also leads to a profound change in the electronic structure. In the recent past, techniques such as aberration corrected scanning transmission electron microscopy (STEM) with combination of annular bright field and high-angle annular dark field imaging \cite{Guzman} and high resolution synchrotron x-ray or neutron scattering including scattering of circularly or linearly polarized x-ray \cite{Walker} were used to track the ion displacements around the phase transition. They were also employed to record the response of the ions under different external stimulation such as electric or magnetic field. The results offer detailed information about the role of ion displacements in inducing different complex functionalities. While large-scale ion displacement of the order of $\sim$10$^{-3}$-10$^{-2}$ \AA is commonly noticed around displacive ferroelectric transition and onset of Jahn-Teller order, movement of ions by comparable scale could also be seen in single-phase multiferroics \cite{Lee} in a regime far away from displacive and/or Jahn-Teller transition. Probing of individual ion movement - which encodes specific information about striction - is expected to offer, in those cases, microscopic picture about the coupling between magnetic and ferroelectric order parameters. Using high resolution powder x-ray and neutron diffraction data - recorded across the magnetic transition temperatures ($T_N$) - we track the anomalous ion displacement patterns in nanoscale ($\approx$20-200 nm) BiFeO$_3$ - the most well-studied room temperature multiferroic \cite{Fiebig}. We find that the unit cell off-centre displacement, magnetoelectric coupling, and the striction driven anomalous displacement of individual ions of the cell around $T_N$ exhibit nonmonotonic pattern of variation with particle size. The particle-size dependence of magnetoelectric coupling has also been probed by direct electrical measurement of remanent ferroelectric hysteresis loops on the nanoparticles under 0-20 kOe magnetic field. 

Extensive research, however, has already been carried out during the last decade on ferroelectric, magnetic, and magnetoelectric properties of BiFeO$_3$ using both isolated nanosized particles and epitaxial thin films. For instance, nearly an order of magnitude improvement of magnetization due to enhanced canting of spins together with anomalous influence of wavelength of the spin spiral ($\sim$62 nm) has been observed in particles of different sizes covering the range 5-100 nm \cite{Park-1,Mazumder,Huang,Landers}. The bandgap, structural noncentrosymmetry as well as soft phonon mode associated with ferroelectric transition have also been probed as a function of particle size \cite{Selbach,Chen}. They appear to be decreasing monotonically with the decrease in particle size. Interestingly, resonant x-ray scattering experiment \cite{Petkov} revealed that Bi sublattice melts below a size limit of $\sim$18 nm. This result contradicts the observation \cite{Chu} of finite ferroelectricity in epitaxial thin films of thickness as small as $\sim$2 nm. In fact, complete 180$^o$ switching of magnetic domains under sweeping electric field has been observed \cite{Heron} in epitaxial thin film of BiFeO$_3$. Epitaxial strain engineering was also shown to induce rhombohedral to orthorhombic phase transition \cite{Yang} and even formation of morphotropic phase boundary \cite{Zeches} with coexisting rhombohedral and tetragonal phases. The influence of microstrain in isolated nanoparticles, of course, has not been probed in detail. More importantly, neither epitaxial thin films nor isolated nanosized particles have been used to generate a comprehensive map of size dependence of magnetoelectric coupling. The complete map of ferroelectric and magnetoelectric properties across a particle size range $\approx$20-200 nm in strained nanoparticles of BiFeO$_3$, being reported here, assumes importance in this backdrop. Attempt has also been made to reconcile the observations of particle-size dependence of off-centred displacement of ions and magnetoelectric coupling from the study of the collective ion movement patterns - determined from group theoretical analysis and refinement of x-ray and neutron diffraction data. While anomalous displacement of Fe ions is found to follow $\tau_1$ mode over the entire particle size range, displacement of O ions appears to undergo a crossover from $\tau_1$ to $\tau_2$ mode as the size reduces below $\sim$30 nm. However, no clear signature of structural phase transition could be noticed in this size range. This is consistent with the observations made by others \cite{Park-1,Mazumder,Huang,Landers,Selbach,Chen,Petkov,Apostolova}.\\

\begin{figure}[ht!]
\centering
\includegraphics[scale=0.40]{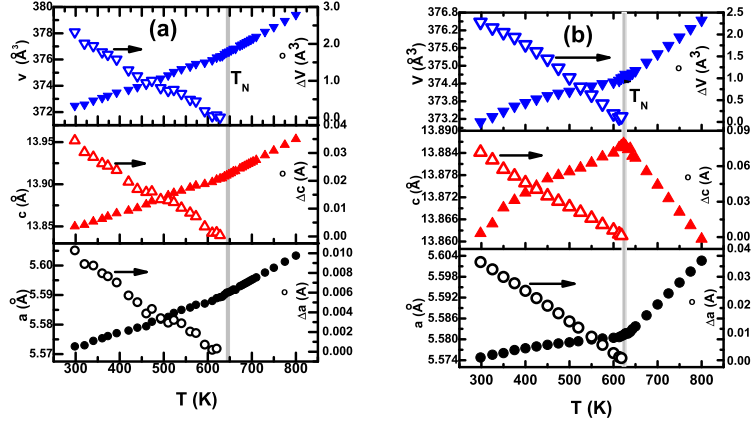}
\caption{(color online) Variation of lattice parameters and volume with temperature for (a) bulk and (b) nanoscale ($\sim$20 nm) BiFeO$_3$; estimated standard deviation obtained from the refinement was used to draw the error bars; they vary within 0.2-0.4\% for the lattice parameters.}
\end{figure}
  
{\Large \noindent \bf Results}

\noindent Figures 1a and 1b show the variation of lattice parameters - $a$ and $c$ - and lattice volume ($v$) with temperature for bulk and nanosized sample ($\sim$20 nm), respectively. Such details for a few other samples are shown in the supplementary document. A clear anomaly around $T_N$ could be noticed for all these parameters signifying presence of spin-lattice coupling. The $T_N$ for all the samples were determined from magnetic and calorimetric measurements. The variation of $T_N$ with particle size is shown in the supplementary document.  The extent of change in the lattice parameters as a result of onset of long-range magnetic order at $T_N$ was determined by using the standard procedure of fitting the data at above $T_N$ and extrapolating the pattern obtained from fitting in the temperature range below $T_N$. This pattern would have been followed by an isostructural yet non-magnetic compound in the absence of any magnetic transition at $T_N$. The subtraction of the actual lattice parameters from the extrapolated ones then yields the extent of change $\Delta a$, $\Delta c$, and $\Delta v$. They are plotted as a function of temperature in Figs. 1a and 1b (right panels). Interestingly, while clear contraction of the lattice parameters and volume could be noticed at $T_N$ in nanoscale samples of intermediate size range (see supplementary document), the bulk sample and finer particles ($\sim$20 nm) do not depict such a feature. This is because the anomaly in the bulk sample is expected to be quite sharp which could not be captured by the diffraction patterns recorded at relatively larger temperature intervals around $T_N$. The sharp feature of the anomaly in bulk sample is reflected in the calorimetric trace (shown in the supplementary document) around $T_N$. The endothermic peak observed at $T_N$ is indeed sharper (full width at half maximum $\approx$2.4 K) than the temperature steps ($\sim$5 K) at which the diffraction patterns have been recorded. The diffraction scans at even smaller temperature steps could not be recorded because of limited beam time availability. In the case of $\sim$20 nm particles, on the other hand, the transition is quite broadened. Therefore, contraction of lattice parameters and volume at $T_N$ could not be noticed in this case too.

\begin{figure}[ht!]
\centering
\includegraphics[scale=0.40]{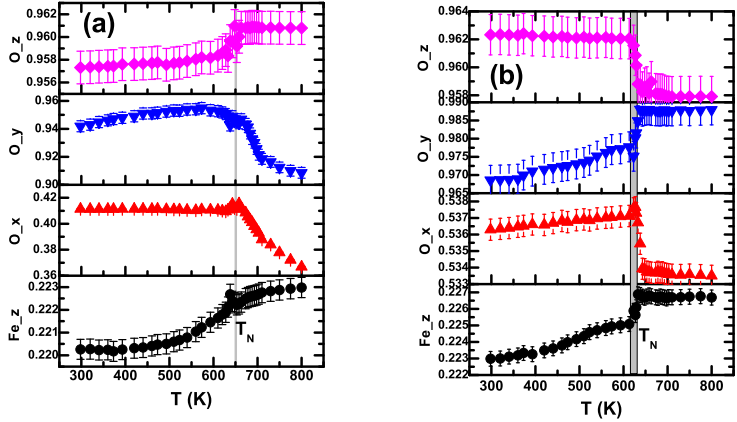}
\caption{(color online) Variation of ion positions with temperature for (a) bulk and (b) nanoscale ($\sim$20 nm) BiFeO$_3$; the estimated standard deviation obtained from the refinement was used for drawing the error bars. }
\end{figure}
 
We now bring out the core component of our work: tracking the movement of individual ions over the entire temperature range 300-800 K and investigating the influence of long-range magnetic order. The results obtained from laboratory and synchrotron x-ray diffraction as well as from neutron diffraction are analyzed and compared in order to establish their intrinsic consistency. The comparison is shown in the supplementary document. It is important to point out here that tracking of individual ion movement around $T_N$ using neutron diffraction data alone poses problem, especially, for the present case where propagation vector k = 0 [Ref. 19]. This is because of appearance of both nuclear and magnetic peaks at the same reciprocal lattice space and difficulty in determining the peak intensity separately for the nuclear and magnetic structure. In order to get around this problem, we compared the x-ray and neutron diffraction data across the entire temperature range to establish the consistency in ion position obtained. Figure 2 shows the Fe 6a (0,0,z) and O 18b (x,y,z) positions (designated for R3c in hexagonal setting) as a function of temperature for the bulk sample and $\sim$20 nm particles. The position of Bi (6a) was kept fixed as the origin (0,0,0). The extent of anomalous movement around $T_N$ turns out to be enormous: for Fe, it is varying within $\sim$0.015-0.113 $\AA$ while for O, the range is $\sim$0.002-0.1 \AA. Such a large scale ion movement has earlier been observed in other multiferroic systems too \cite{Lee}. In fact, large scale movement of ions near $T_N$ by the extent expected in displacive ferroelectric systems around the ferroelectric transition was considered a plausible microscopic mechanism behind coupling between magnetic and ferroelectric order parameters in multiferroic systems \cite{Lee}. In the present case, we used anomalous movement of Fe and O ions below $T_N$ for noting the striction on individual ions. We plot the variation of anomalous ion displacement at $T_N$ as a function of particle size in Fig. 3. Surprisingly, the displacement of the ions at $T_N$ appears to vary nonmonotonically. In spite of the fact that the transition at $T_N$ appears to be isostructural in all the cases (as no clear signature of structural phase transition could be noticed) and magnetization increases monotonically with the decrease in particle size \cite{Park-1,Mazumder}, movement of individual ions below $T_N$ does not seem to be governed by magnetization alone. A different mechanism is also at play to reduce the anomalous ion movement around $T_N$ in finer particles even in presence of large magnetization. 

\begin{figure}[ht!]
\centering
\includegraphics[scale=0.25]{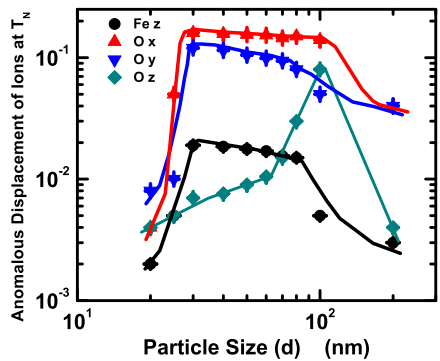}
\caption{(color online) Particle size dependence of anomalous displacement of Fe and O ions at $T_N$.}
\end{figure}

\begin{figure}[ht!]
\centering
\includegraphics[scale=0.25]{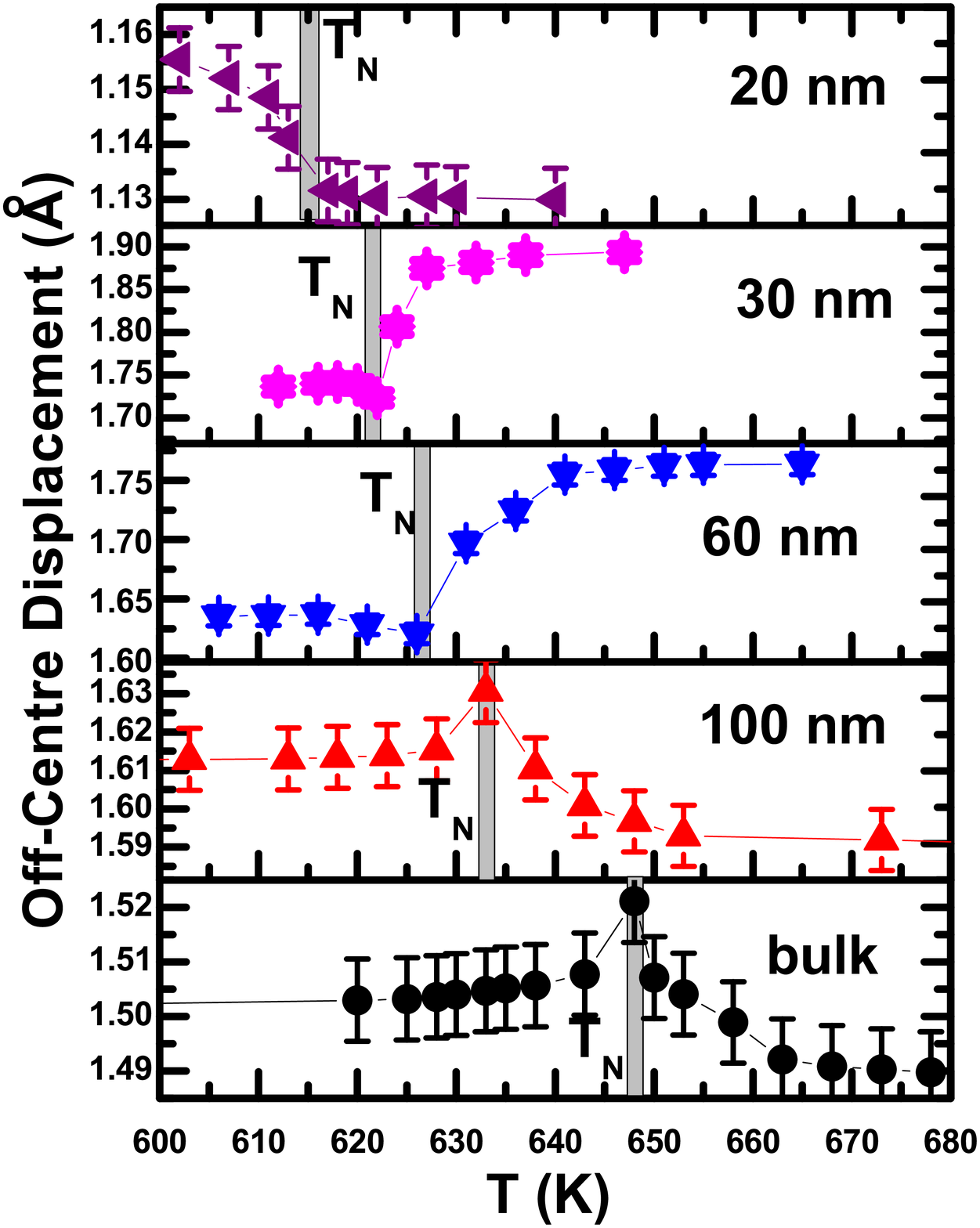}
\caption{(color online) Variation of off-centre displacement within a unit cell with temperature across $T_N$ for a few representative samples. }
\end{figure}

\begin{figure}[ht!]
\centering
\includegraphics[scale=0.25]{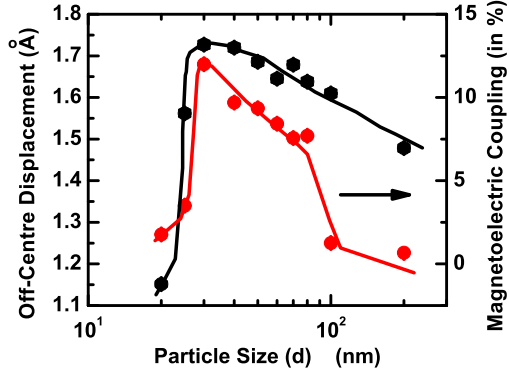}
\caption{(color online) Particle size dependence of off-centre displacement within a cell and magnetoelectric coupling.}
\end{figure}

We next investigate the off-centering of the ions within a unit cell and its variation with particle size. We first determined the position of the ions in a unit cell from the refinement of the diffraction data. In hexagonal setting, the axis of polarization [111] for BiFeO$_3$ (space group $R3c$) transforms to [001]. The off-centered displacement of Bi and Fe ions along [001] from their centrosymmetric positions with respect to the near-neighbour oxygen ions has been determined using the respective position of the ions. The net off-centering of Bi and Fe ions along [001] is then used to determine the overall off-centered displacement in a unit cell ($\delta$). This procedure has earlier been followed by Selbach $\textit{et al}$. \cite{Selbach}. 

Finally, we turn our attention to the issue of variation of magnetoelectric coupling with particle size. The magnetoelectric coupling of intrinsic multiferroic origin is reflected in the anomaly in $\delta$ around $T_N$. The $\delta$ in a cell - calculated as mentioned above - is plotted as a function of temperature ($T$) [Fig. 4]. An anomaly around the $T_N$ is conspicuous in all the cases. The $\delta$ has either decreased or enhanced as the temperature is reduced through $T_N$ from above. In the case of bulk sample as well as coarser particles, where the $\delta$ is relatively smaller, it exhibits a rise below $T_N$. In finer particles, where $\delta$ is larger, it shows a decrease below $T_N$. In even finer particles, where $\delta$ becomes smaller again, it rises below $T_N$. Of course, while the gross feature of rise in $\delta$ with the drop in temperature from far above $T_N$ to far below $T_N$ prevails both in bulk sample and coarser particles as well as in $\sim$20 nm particles, a small peak-like feature at $T_N$ is also associated in the case of bulk sample and coarser particles. From neutron diffraction experiment on a single crystal of BiFeO$_3$, it has been shown earlier \cite{Park-2} that the change in polarization (or off-centre displacement of ions) below $T_N$ results primarily from magnetostrictive effect which leads to its decrease and not from inverse Dzyloshinskii-Moriya (DM) effect which is supposed to increase it. In the present case, however, we observe both increase and decrease in $\delta$ below $T_N$ even though, as pointed out later, this change is primarily governed by the striction effect. The extent of anomaly - as this will give a quantitative estimate of the extent of multiferroic coupling in bulk and nanoscale BiFeO$_3$ - is calculated from $\frac{\delta_{av (below T_{N})}-\delta_{av (above T_{N})}}{\delta_{av (above T_{N})}}$. The data are used to arrive at our prime result on the variation of magnetoelectric multiferroic coupling with particle size in bulk and nanoscale BiFeO$_3$ (Fig. 5). It appears that the coupling varies nonmonotonically (within $\sim$0.5\%-12\%) with the particle size for the range covered in this study. There seems to exist an optimum particle size ($\sim$30 nm) at which the coupling maximizes. Comparison of the results obtained from both x-ray and neutron diffraction data exhibits an intrinsic consistency in the magnitude of the magnetoelectric coupling parameter. For example, magnetoelectric coupling turns out to be $\sim$1\% for the bulk sample in both the x-ray and neutron diffraction experiments whereas for the $\sim$30 nm particles the coupling is found to be $\sim$12\% and $\sim$13.5\%, respectively. Figure 5 also shows the variation of net off-centre displacement of the ions ($\delta$) at room temperature as a function of particle size. Interestingly, this plot too is nonmonotonic. The off-centre displacement increases initially as the particle size decreases and then below an optimum size ($\sim$30 nm), it drops off.    

In order to verify further the observation of nonmonotonic variation of magnetoelectric coupling with particle size, we have carried out direct electrical measurements, as well, on a few selected samples with different nanosized particles acoss the entire range $\approx$20-200 nm. The particles were deposited on an insulating substrate (Si/SiO$_2$) in the form of a film and electrodes (Cu/Ag) were deposited in two-probe configuration. The schematic of the sample-electrode configuration is shown in the supplementary document. Similar sample-electrode configuration was earlier used by others \cite{Sun-1,Sun-2} for measuring the dielectric and ferroelectric properties of nanoscale particles. For a limited case, we have also used sophisticated e-beam lithography and focused ion beam for patterning the electrodes on nanochains of BiFeO$_3$ as described in one of our earlier papers \cite{Goswami-1}. We have measured the remanent ferroelectric polarization ($P_R$) of the samples under different magnetic fields across 0-20 kOe in order to determine the change in the polarization under magnetic field. This change - given in \% by $\frac{P_R(0)-P_R(20 kOe)}{P_R(0)}$ $\times$ 100 - has been compared with the change observed in off-centered displacement of the ions around the $T_N$. The change in the polarization under magnetic field obtained from the measurement has been used to map the particle size dependence of the magnetoelectric coupling. The remannet ferroelectric hysteresis loops under zero and $\sim$20 kOe magnetic field for a few selected cases as well as the plot of variation in magnetoelectric coupling with particle size - obtained from this direct electrical measurements on the nanoparticles - are shown in the Fig. 6. It is clear from this plot that the gross feature of nonmonotonic variation of magnetoelectric coupling with particle size is corroborated both by diffraction and direct electrical measurements. However, the magnitude of the magnetoelectric coupling differs. This could be because of influence of defects at the domain boundaries which possibly induce variation in the domain switching characteristics in presence and absence of magnetic field.  

\begin{figure*}[ht!]
\centering
\includegraphics[scale=0.35]{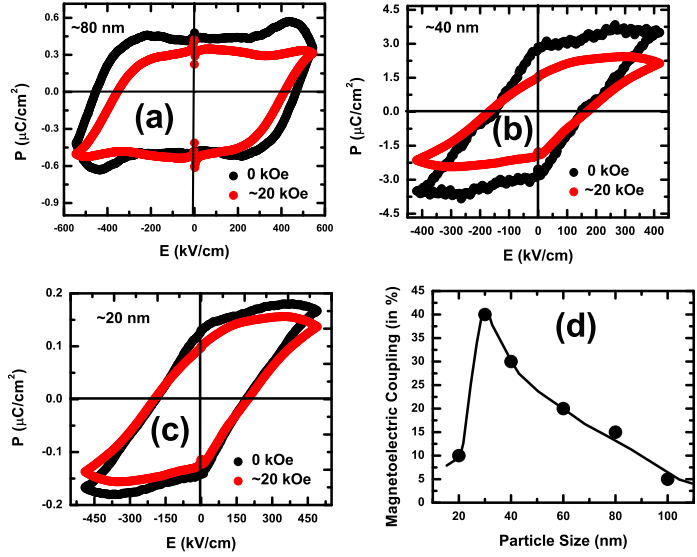}
\caption{The remanent ferroelectric hysteresis loops measured at room temperature under zero and $\sim$20 kOe field for samples containing particles of sizes (a) $\sim$80 nm, (b) $\sim$40 nm, and (c) $\sim$20 nm; (d) the overall pattern of variation of magnetoelectric coupling (in \%) with particle size obtained from this direct electrical measurements on the nanoparticles.}
\end{figure*}  

It is important to point out that we have used a special protocol for extracting the remanent ferroelectric polarization of the samples. This protocol sends out fourteen pulses (square prepolarization pulses and triangular polarization and measurement pulses) to the sample in modified Sawyer-Tower circuit to measure, separately, the contribution of ferroelectric and nonferroelectric polarization as well as the contribution of nonferroelectric polarization alone. Subtraction of the hysteresis loop obtained for the second case from the former one yields the intrinsic switchable remanent ferroelectric polarization. The details of the protocol, how it works, and the underlying physics have been described in our earlier work \cite{Chowdhury}. We ensure that intrinsic remanent ferroelectric polarization is indeed obtained from this protocol for the samples used in present case.

The rise in magnetoelectric coupling in nanoscale BiFeO$_3$ is expected as magnetization enhances in finer particles due to incomplete spin spiral and enhanced spin canting \cite{Park-1,Mazumder}. From the point of view of crystallographic structure, magnetization is contained in (111) plane. With the increase in magnetization, antiferrodistortive rotation of FeO$_6$ octahedra around the polarization axis [111] too enhances \cite{Ederer}. This, in turn, influences the polarization by changing the polar displacement of the ions along the [111]. However, with further decrease in particle size below $\sim$30 nm, compressive pressure appears to govern the properties.

Comparison of the results shown in Figs. 3 and 5 yields an interesting trend. While magnetization increase monotonically with the decrease in particle size \cite{Park-1,Mazumder}, the striction driven anomalous displacement of an ion, the net off-centred displacement of the ions, and, finally, the magnetoelectric coupling exhibit nonmonotonicity. In order to reconcile the particle size dependence of all these properties we analyzed the anomalous movement patterns of individual ions around the $T_N$ using group theoretical approach within the framework of isostructural transition. This group theoretical analysis of the anomalous ion displacement modes around $T_N$ assumes importance from the point of view of understanding the nature of the transition - isostructural or symmetry-changing - as well. We used $\textit{BasIreps}$ within the FullProf Suite platform to determine the irreducible representations ($\tau_1$, $\tau_2$, and $\tau_3$) which define all the allowed collective ion displacement patterns for Fe (site 6a) and O (site 18b) ions at $T_N$ for $R3c$ symmetry with propagation vector k = 0. We follow the notations used in Ref. 5 for designating the irreducible representations. The basis functions for all the irreducible representations are given in the supplementary document. Using the basis functions corresponding to the displacement modes $\tau_1$ and $\tau_2$, the possible displacement patterns of Fe (6a) and O (18b) ions are also shown in the supplementary document. By comparing this theoretical result with the experimentally observed displacement patterns obtained from refinement of x-ray and neutron diffraction data, we discover that the anomalous displacement of Fe ions around $T_N$ is actually consistent with $\tau_1$ mode throughout the entire particle size range. O ion movements, on the other hand, exhibit a transformation from $\tau_1$ to $\tau_2$ mode in particles finer than $\sim$30 nm. However, the transition at $T_N$ turns out to be isostructural for the entire particle size range ($\approx$20-200 nm). No signature of structural transition at $T_N$ could be noticed even for finer particles where O ions exhibit $\tau_2$ mode of anomalous displacement. This could be because the crystal class is preserved even for $\tau_2$ mode of displacement. Others also did not report \cite{Park-1,Huang,Landers,Selbach,Petkov} any structural phase transition in nanoscale BiFeO$_3$ within this size range. The isostructural transition is a bit rare \cite{Scott}. The physics behind isostructural transition is not quite well understood though the electron-lattice interaction was considered important in this context \cite{Barma}. The change in symmetry as a result of subtle displacement of ions in finer particles, if at all has taken place, could not be captured in the present case. The experimentally observed cooperative ion movement patterns for bulk sample and $\sim$20 nm particles are shown in Fig. 6 using $R3c$ structure. While displacement of Fe ions is confined along c-axis, O ion displacement takes along a,b, and c-axes. However, for clarity, only the D$_1^1$ part of the displacement for O ions are shown in Fig. 7. For this part, O ion movement is restricted within ab-plane - either along a- or b-axes or at an angle of 45$^o$ with the axes. The basis functions corresponding to the movement direction are shown in Fig. 7 for O ions. 

\begin{figure}[ht!]
\centering
\includegraphics[scale=0.65]{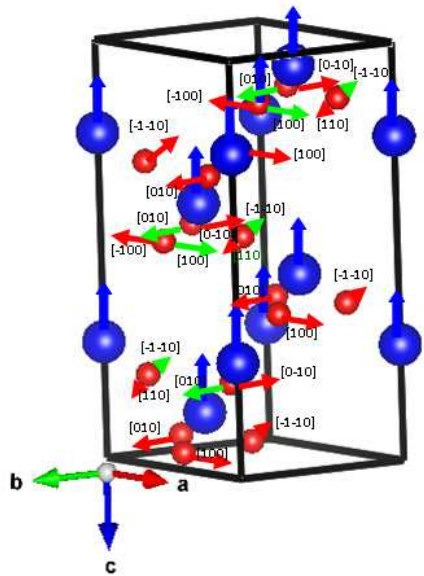}
\caption{(color online) The Fe (blue) and O (red) ion displacements (the D$_1$ part) observed in bulk sample and $\sim$20 nm particles are shown. Fe ion displacements are shown by blue arrows while O ion displacements are shown by red and green arrows. The green arrows show the $\tau_2$ mode of displacement while other arrows show the $\tau_1$ mode. For clarity, Bi ions are not shown here.}
\end{figure}

\begin{table*}[ht]
{\small
\caption{The strain, lattice parameters, ion positions, and compressive pressure as obtained from first principle calculations using density functional theory. The ion positions obtained are used for calculating the off-centre displacement in a cell. The results simulate the experimentally observed nonmonotonic particle-size-dependence of off-centre displacement in nanoscale BiFeO$_3$.}

\begin{tabular}{p{0.3in}p{0.3in}p{0.4in}p{1.4in}p{1.4in}p{1.4in}} \hline\hline
Strain \newline (\%) & $a$ \newline (\AA) & $c$ \newline (\AA) & Bi \newline x, y, z & Fe \newline x, y, z & O \newline x, y, z \\ \hline 
\\
0 & 5.446 & 13.144 & 0.0000 0.0000 -0.01244 & 0.0000 0.0000 0.21891 & 0.24436 0.89577 0.29104\\
\\
-1.0 & 5.529 & 13.241 & 0.0000 0.0000 -0.00959 & 0.0000 0.0000 0.21896 & 0.24625 0.89726 0.29008\\
\\
-2.0 & 5.500 & 13.472 & 0.0000 0.0000 0.00750 & 0.0000 0.0000 0.78156 & 0.01478 0.43679 0.04378\\
\\
+1.0 & 5.539 & 12.825 & 0.0000 0.0000 -0.01461 & 0.0000 0.0000 0.21997 & 0.24855 0.89875 0.29142\\
\\
+2.0 & 5.601 & 12.540 & 0.0000 0.0000 -0.01716 & 0.0000 0.0000 0.22091 & 0.25167 -0.09918 0.29195\\ 
\\ \hline 
\\
-2.0 \newline (0.063 GPa) & 5.500 & 13.2156 & 0.0000 0.0000 -0.01656 & 0.0000 0.0000 0.22147 & 0.25116 -0.09957 0.29257\\
\\
-2.0 \newline (0.7 Gpa) & 5.500 & 13.0542 & 0.0000 0.0000 -0.01659 & 0.0000 0.0000 0.22114 & 0.25082 -0.09975 0.2926\\
\\
-2.0 \newline (7.7 GPa) & 5.500 & 12.8468 & 0.0000 0.0000 -0.01725 & 0.0000 0.0000 0.22037 & 0.24908 0.8993 0.29216\\ \hline \hline 

\end{tabular}}

\end{table*}

\begin{figure}[ht!]
\centering
\includegraphics[scale=0.30]{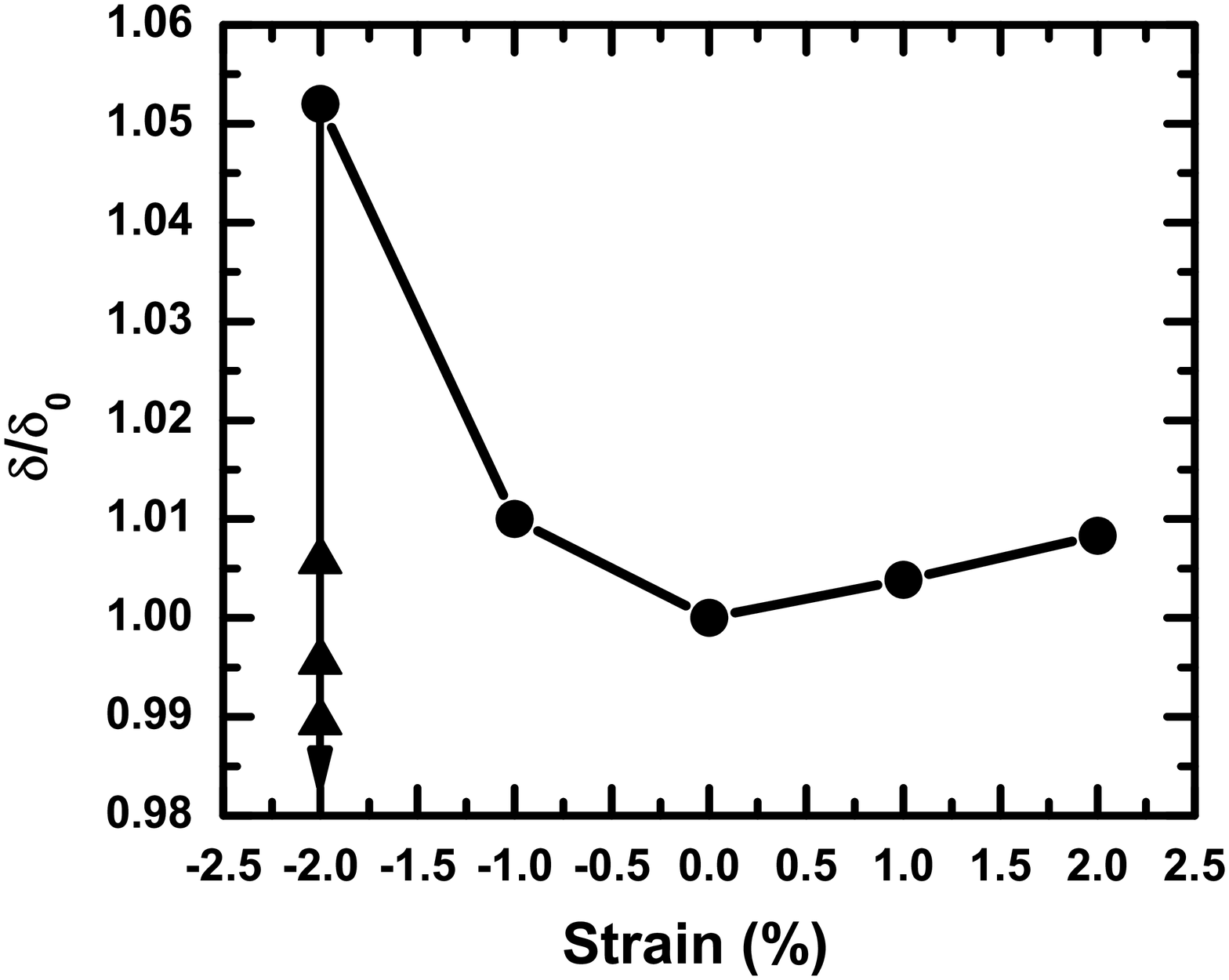}
\caption{The variation of off-centre displacement ($\delta$), normalized by the value ($\delta_0$) for strain-free sample, with lattice strain (circles) as obtained from first principle calculations; the results obtained under compressive pressures in presence of large lattice strain are also shown (triangles).}
\end{figure}

The origin of such anomalous displacement patterns of Fe and O ions at respective $T_N$s and consequent nonmonotonicity in particle-size-dependence of different properties has been investigated further. The Rietveld refinement of the diffraction patterns reveals a monotonic rise in lattice strain. This plot is given in the supplementray document. It appears that these as-prepared particles contain lattice strain and the strain enhances as the size decreases. No special annealing treatment has been employed to release the strain. With the decrease in particle size, an internal pressure ($\bar{P}$) too increases which is given approximately by the Young-Laplace equation $\bar{P}$ = 2$S$/$d$, where $S$ is the surface tension (of the order of 50 N/m for the perovskite oxides \cite{Zhou}) and $d$ is the particle size. While large lattice strain, via its coupling with ferroelectric order, enhances the polarization, increased compressive pressure reduces it as pressure drives the crystallographic structure toward centrosymmetry. In fact, it has been shown that with the increase in pressure the rhombohedral BiFeO$_3$ undergoes a series of structural phase transition to centrosymmetric structures \cite{Houmont,Guennou}. The first transition from $R3c$ to centrosymmetric $C2/m$ could be noticed at a pressure as small as $\sim$3 GPa. In the present case, according to the relation mentioned above, an internal pressure of $\sim$5.0 GPa could be generated in particles of size $\sim$20 nm. The progressive suppression of noncentrosymmetry observed in a bulk sample under directly applied external pressure and similar observation in nanoscale particles with the decrease in particle size offers an additional evidence that indeed pressure builds up in nanosized particles. No structural transition could, however, be observed. This could be because this estimate is just approximate and cannot be directly equated with the observations made under externally applied compressive pressure. Of course, the competition between enhanced pressure and consequent drop in noncentrosymmetry and the strain-driven rise in off-centering seems to be the origin of the observed nonmonotonic pattern of net $\delta$. This is further reflected in the plot of lattice volume versus particle size included in the supplementary document. The lattice volume increases with the decrease in particle size within $\sim$30 nm to $\sim$200 nm range and then decreases below $\sim$30 nm. The observed crossover in the collective displacement patterns of oxygen ions from $\tau_1$ to $\tau_2$ mode in finer particles could also be due to the competition between strain and pressure effects. We used first principle density functional theory to calculate the structural noncentrosymmetry in presence of competition between enhanced lattice strain and compressive pressure. The plane-wave pseudopotential method was used for the calculations with 5 valence electrons for Bi (6$s^2$, 6$p^3$), 8 for Fe (3$d^6$, 4$s^2$), and 6 for O (2$s^2$, 2$p^4$) as implemented in the $\textit{Vienna Ab initio Simulation Package}$ (VASP). Geometry optimizations were carried out using generalized gradient approximation (GGA) within the framework of Perdew - Burke - Ernzerhof (PBE) for the exchange and correlation functional. The self-consistent convergence accuracy and cut-off energy were set at 10$^{-6}$ eV/atom and 400 eV respectively. Highly dense 21 $\times$ 21 $\times$ 21 Monkhorst - Pack k-point grid was used to sample the Brillouin zone, while tetrahedron method with Bl$\ddot{o}$chl correction was used for Brillouin zone integration. Convergence criterion for the maximal Hellman - Feynman force between atoms and Gaussian Broadening were set at 0.01 eV \AA$^{-1}$ and 0.05 eV during relaxation of the ions. Initially, the influence of lattice strain (both compressive and tensile) on noncentrosymmetry under ambient pressure is verified. In order to calculate the effect of lattice strain, lattice parameters for strain-free bulk sample were used. The in-plane parameter $a$ was constrained depending on the extent of strain introduced while the out-of-plane parameter $c$ and all the ion positions were varied. The relaxed lattice parameter $c$ and the ion positions as well as electronic structure were finally determined for the minimum total energy state under such condition. This calculation was then repeated for different strain. Although this protocol actually yields the impact of epitaxial strain, because of similarity in the gross features \cite{Mocherla,Tajiri,Satar,Martin,Pratt}, it, as well, captures the essence of influence of microstrain in a nanosized particle. In spite of differences such as association of disorder with microstrain, it appears that microstrain too influences the physical properties such as ferroelectricity, magnetism, charge transport etc in the same way as epitaxial strain. The similarity between the effect of epitaxial and microstrain is illustrated clearly in the study of variation of bandgap in nanoparticles of BiFeO$_3$ as a function of strain \cite{Satar}. The pattern observed is comparable to similar pattern in epitaxial thin films containing epitaxial strain. Therefore, even though this is a toy model, used for examining the variation of structural noncentrosymmetry in presence of competition between lattice strain and compressive pressure, the results obtained from the model are relevant in the context of microstrain in nanoparticles. Direct simulation of the entire nanoparticle of $\approx$20-200 nm size which requires consideration of large number of atoms is beyond the scope of this work. In Table 1, we show the extent of strain, the lattice parameters, and the ion positions. After determining the map of the strain versus structural parameters, the calculation was carried out for determining the parameters in presence of compressive pressure using the results obtained under maximum strain. The pressure was generated by internal force matrix acting on individual ions. At each level, the ions were allowed to relax. Calculation of the effect of compressive pressure in presence of large lattice strain is the new aspect. This has been done in order to capture the essential physics behind the nonmonotonic particle size dependence of noncentrosymmetry. The lattice parameters and ion positions under pressure are given in the Table 1. The variation of the off-centre displacement with lattice strain and compressive pressures in presence of strain are shown in Fig. 8. It is found that the structural noncentrosymmetry increases by $\approx$5-6\% with the increase in lattice strain across $\pm$2\%. Compressive strain is found to have stronger influence on the noncentrosymmetry. Similar result has earlier been reported for BiFeO$_3$ \cite{Spaldin}. While lattice strain indeed enhances the structural noncentrosymmetry, compressive pressure reduces it progressively and, at a pressure $\sim$7.7 GPa, it gives rise to even smaller noncentrosymmetry than what is observed in a strain-free sample while maintaining the $R3c$ symmetry throughout (Fig. 8). This result is consistent with our experimental observation and explains the physics behind nonmonotonic particle-size-dependence of off-centre displacement. The issue of structural phase transition under compressive pressure within the range 0-7.7 GPa was addressed by calculating the minimum total energy for $R3c$ and possible other phases such as centrosymmetric $C2/m$ and $Pnma$. The $R3c$ appears to offer the lowest minimum total energy. The difference between the minimum total energy for $R3c$ and $C2/m$ phase turns out to be of the order of $\sim$40 eV. The samples exhibiting $R3c$ to $C2/m$ transition \cite{Houmont,Guennou} at a smaller pressure of $\sim$3 GPa may contain large scale lattice defects. In our nanoscale samples too, we did not observe any signature of structural phase transition in finer particles where, as shown above, the compressive pressure could be $\sim$5 GPa.\\

{\Large \noindent \bf Discussion}

\noindent As the structural noncentrosymmetry - measured by off-centering of the ions with respect to their centrosymmetric position at above the ferroelectric transition temperature - is directly related to the ferroelectric polarization, the ion positions and their shift is also closely related to the magnetic spin structure of the lattice. This is because the exchange coupling mechanism which is responsible for giving rise to the long-range spin order depends on the ion positions and distance (i.e., bond length) between the ions. The influence of spin structure on the ion position could be clearly observed in the magnetoelastic effect where spin-lattice coupling yields an anomalous change in the lattice parameters, ion positions, bond lengths and angle. In the present case, we show that onset of long-range magnetic order influences the ion positions to give rise to the change in the net off-centering as well. Therefore, apart from the magnetoelastic effect, one observes magnetoelectric effect as well. This observation is equivalent to the observation of change in polarization under magnetic field at below the $T_N$. It is important to point out, in this context, that in one of our earlier work \cite{Goswami-2} we carried out powder neutron diffraction experiments at room temperature (i.e., well below $T_N$) under $\sim$50 kOe magnetic field. The ion positions as well as the net structural noncentrosymmetry indeed exhibited sizable change under $\sim$50 kOe field reflecting both magnetoleastic and magnetoelectric effects in nanoscale BiFeO$_3$. The close correlation among off-centre displacement of a cell, striction of an ion around $T_N$, and magnetoelectric coupling indicates presence of strong coupling between piezo and magnetostriction. Via this coupling, ferroelectric polarization appears to be influencing the striction of an ion directly which, in turn, governs the magnetoelectric coupling. The clear dependence of magnetoelectric coupling on the net off-centre displacement rules out inverse DM mechanism to be the origin of magnetoelectric coupling. The influence of piezostriction is more apparent in finer particles where in spite of rise in magnetization, striction driven anomalous ion movement at $T_N$ exhibits a drop following the trend of polarization. Therefore, while rise in lattice strain enhances the structural noncentrosymmetry in finer particles, rise in compressive pressure tends to reduce it. The competition yields nonmonotonic particle-size-dependence of noncentrosymmetry as well as magnetoelectric coupling. This pattern is driven by nonmonotonic variation of striction of an individual ion around $T_N$ with particle size.  

In summary, we offer a comprehensive map of particle size dependence of multiferroic properties across a wide range of particle size - from bulk ($\sim$200 nm grain size) to nanoscale ($\sim$20 nm) - for BiFeO$_3$. We show that with the decrease in particle size, the structural noncentrosymmetry, magnetoelectric coupling, and anomalous displacement of individual ions at $T_N$ follow a nonmonotonic pattern. This behavior is consistent with change in anomalous ion displacement patterns around $T_N$ in finer particles. The nonmonotonicity results from competition between enhanced compressive pressure and lattice strain in finer particles. Observation of maximum polarization and magnetoelectric coupling in $\sim$30 nm particles could have profound effect on designing nano-spintronic devices based on nanoscale BiFeO$_3$.\\  

{\Large \noindent \bf Methods}

\noindent \small{The BiFeO$_3$ nanoparticles of different sizes were prepared by sonochemical route. The details of the sample preparation have been described elsewhere \cite{Goswami-2,Goswami-3}. For this work, samples of average particle size 20, 25, 30, 40, 50, 60, 70, 80, 100, and 200 nm were used. The particle size, its distribution, shape, morphology, crystallographic structure etc were studied by transmission electron microscopy (TEM). Representative results for particles of average size $\sim$20, $\sim$30, and $\sim$60 nm are shown in the supplementary document. The magnetic transition temperature ($T_N$) were determined from calorimetry and magnetic measurements carried out across 300-800 K. The $T_N$ decreases from $\sim$653 K in bulk to $\sim$615 K in particles of size $\sim$20 nm. This is consistent with observations made by others \cite{Selbach}. Because of incomplete spin spiral in particles of size below $\sim$62 nm (the wavelength of the spiral) and increase in spin canting angle, the antiferromagnetic order weakens, leading to a drop in the $T_N$ in finer particles. High resolution powder x-ray diffraction patterns were recorded at different temperatures across $T_N$ (within 300-800 K range) by laboratory diffractometer (Bruker D8 Advance) as well as (for a few cases) synchrotron radiation facility at ESRF (beamline ID22; $\lambda$ = 0.3191 \AA) and Photon Factory, Tsukuba, Japan (beamline BL-18B; $\lambda$ = 0.88 \AA). Powder neutron diffraction data have also been collected (at PD-2 diffractometer of NFNBR, Mumbai, $\lambda$ = 1.2443 \AA; PD-3 diffractometer, $\lambda$ = 1.48 \AA) both under zero magnetic field across the respective $T_N$ of a sample and under a field of $\sim$50 kOe at room temperature (i.e., at well below $T_N$). The diffraction data were refined by FullProf in order to extract the structural details such as space group, lattice parameters, ion positions, bond lengths, angles, distortion, magnetic moment, crystallite size, microstrain in the lattice etc for all the samples. The supplementary document contains these details. The microstrain has been calculated during refinement by turning on the microstructural properties using the standard procedure within Fullprof \cite{Fullprof}. The $R_p$, $R_{wp}$, and $\chi^2$, given in the supplementary document, turn out to be varying within 5\%-25\% and 2.0-8.0. This is within the acceptable range. The estimated standard deviation has been used for drawing the error bars. The space group turns out to be R3c in all the cases both at above and below $T_N$. The magnetoelectric coupling and magnetoelastic effect as a function of particle size have been determined from the structural details obtained from refinement. For a few selected samples, direct electrical measurements have been carried out in order to determine the change in remanent ferroelectric polarization under a magnetic field. The schematic of the two-probe electrode configuration used on a film of the nanoparticles is shown in the supplementary document. The measurement of remanent hysteresis loop under different magnetic fields across 0-20 kOe has been carried out by the ferroelectric loop tester of Radiant Inc. (Precision LCII).}

\vspace{10mm}
 
\noindent {\Large \textbf{Acknowledgements}}\

\noindent One of the authors (S.G.) acknowledges support in the form of Senior Research Associateship from CSIR, Government of India, during this work.\\

\noindent The supplementary information are available on request.


\begin{thebibliography}  {99}
\bibitem{footnote} Even though the magnetic structure has a spatial modulation over a length scale of $\sim$62 nm, it is found that the magnetic lattice can be described quite well for bulk as well as nanoscale samples by using propagation vector k = 0.
\end{thebibliography}
\end{document}